\documentclass[times,twocolumn,final]{elsarticle}

\usepackage{journal}

\usepackage{framed,multirow}

\usepackage{amssymb}
\usepackage{latexsym}

\usepackage{url}
\usepackage{xcolor}
\usepackage{amsmath}
\usepackage{amssymb}
\usepackage{color}
\usepackage{caption}
\usepackage{multirow}
\usepackage{booktabs} 
\usepackage{graphicx}
\usepackage{latexsym}
\usepackage[version=4]{mhchem}
\usepackage{hyperref}
\usepackage[normalem]{ulem} 
\usepackage{rotating}
\definecolor{newcolor}{rgb}{.8,.349,.1}
\usepackage{colortbl}
\usepackage{hyperref}
\usepackage[table]{xcolor} 
\usepackage{siunitx}
\usepackage[switch,pagewise]{lineno} 

\journal{}

\newcommand{\ghr}[0]{$\mathcal{I}_{HR}$}
\newcommand{\glr}[0]{$\mathcal{I}_{LR}$}
\newcommand{\gint}[0]{$\mathcal{I}_{Int}$}
\newcommand{\gsr}[0]{$\mathcal{I}_{SR}$}
\newcommand{\gcnn}[0]{$\mathcal{I}_{CNN}$}

\newcommand{\bs}[0]{\boldsymbol}

\begin{document}

\verso{Dash et al.}

\begin{frontmatter}

\title{Super-resolution of turbulent reacting flows on complex meshes using graph neural networks}%

\author[1]{Priyabrat \snm{Dash}\corref{cor1}}
\cortext[cor1]{Corresponding author}
\emailauthor{priyabratd@iisc.ac.in}{Priyabrat Dash}
\author[1]{Konduri \snm{Aditya}}
\author[2]{Christos E. \snm{Frouzakis}}
\author[3]{Mathis \snm{Bode}}
\address[1]{FLAME Laboratory, Department of Computational and Data Sciences, Indian Institute of Science, Bengaluru, 560012, India}
\address[2]{CAPS Laboratory, Department of Mechanical and Process Engineering, ETH Zürich, 8092 Zürich, Switzerland}
\address[3]{Jülich Supercomputing Centre, Forschungszentrum Jülich GmbH, Jülich, 52425, Germany}

\begin{abstract}
State-of-the-art deep learning models have been extensively utilized to reconstruct small-scale structures from coarse-grained data in turbulent flows. However, their application has predominantly been restricted to structured uniform meshes, limiting their applicability to data associated with complex geometries that are typically simulated on structured non-uniform or unstructured meshes. Machine learning (ML) models based on graph neural networks (GNNs), known for their ability to process unstructured data, offer a promising alternative. In this study, we leverage the inherent flexibility of GNNs featuring message passing layers to develop a methodology for reconstructing unresolved small-scale structures from low-resolution data on complex meshes. The accuracy of the proposed approach is demonstrated using two cases: a reacting channel flow on a structured non-uniform mesh, and a reacting hydrogen fueled internal combustion {(IC)} engine featuring an unstructured mesh. Evaluation of results based on visual agreement, statistical metrics, and cumulative error reduction indicates the effectiveness of the method in accurately reconstructing fine-scale features. Overall, this study provides a pathway for integrating data-driven small-scale reconstruction and subgrid-scale modeling to enhance the accuracy of coarse-grained simulations on complex meshes.
\end{abstract}


\end{frontmatter}


\section*{Novelty and significance statement}
{This work presents a novel graph neural network (GNN) framework with message passing layers for interpolation-free super resolution of three-dimensional turbulent reacting flow data on complex meshes. Rigorous evaluation through gradient profiles and joint probability density functions confirms that the framework provides the reconstruction accuracy required for high-fidelity subgrid scale modeling. By handling simulation data natively on complex meshes relevant to practical combustors such as internal combustion engines, this method avoids resampling inaccuracies to preserve essential fine-scale features. Consequently, this approach contributes to advancing data-driven models for refining simulation and experimental data to facilitate better extraction of physical insights.}

\section{Introduction\label{sec:introduction}} 

Turbulence exhibits complex, chaotic structures spanning a wide range of spatio-temporal scales, posing challenges for predictive simulations. Direct numerical simulation (DNS) addresses this challenge by resolving all dynamically relevant turbulent scales. Such calculations become computationally prohibitive at high Reynolds numbers, in geometrically complex domains, and for reacting flows. This motivates approaches for recovering unresolved flow structures from relatively inexpensive simulations. Artificial intelligence (AI) based super-resolution has proven to be a useful tool in many scientific fields, ranging from image processing \citep{Wang2019} to astrophysics \citep{LiYin2021}. The general idea is to use an artificial neural network trained on high-fidelity data to supplement missing features in coarse data. This also results in many direct applications in the field of fluid mechanics, especially in turbulent flows and combustion: noisy experimental data can be quantitatively improved \cite{Yu2024}, non-intrusive alternatives of inherently intrusive measurement techniques can be developed \cite{Dash2024TVC}, simulations based on reduced order models (ROMs) (specifically, large-eddy simulation (LES)), can be locally refined to model or understand physical effects that are strongly influenced by the smallest scales \cite{Bode2022NP, Bode2023Chapter}, initial conditions and turbulent inflow \cite{Liu2025}  can be conveniently generated, and the complete resolution of flame fronts becomes possible \cite{Liu2024}. The wide applicability of this approach motivates a more detailed examination of existing shortcomings, such as problems with 3D reconstruction, the systematic and accurate handling of flame fronts, and technical challenges such as complex meshes. This paper is a  step in this direction, demonstrating an accurate (interpolation free) application of machine learning based super-resolution to reacting flows on complex meshes using a model based on graph neural networks (GNNs) \citep{Battaglia2018}.

Current advances in super-resolution models for fluid dynamics span various deep learning techniques, including convolutional neural networks (CNNs) \cite{Fukami2024}, generative adversarial networks (GANs) \cite{Subramaniam2020}, physics-informed neural networks (PINNs) \cite{Li2021FNO}, transformers \cite{Li2024}, and denoising diffusion probabilistic models (DDPMs) \cite{Shu2023}. These approaches have shown considerable promise in improving the resolution of fluid simulations across different scenarios. CNNs trained with pixel-wise loss are particularly effective at capturing local spatial features, excelling in reconstructing turbulent flow data. Beyond CNNs, GANs have garnered attention for their capacity to generate highly realistic high-resolution outputs. However, while visually compelling, such models often fail to maintain physical consistency in fluid behavior, as they primarily optimize for statistical similarity to training data rather than the underlying physical laws. PINNs \cite{Li2021FNO} address this limitation by integrating physical equations into their architecture, ensuring that the generated high-resolution simulations adhere to fundamental fluid mechanics principles. Despite their promise, PINNs can struggle with highly turbulent flows where simplified equations fall short of capturing the full complexity of fluid interactions, particularly in cases involving multiple interacting scales. Physics-informed extensions of enhanced super-resolution GANs (PIESRGANs) utilize a combination of adversarial and physics-based losses for subgrid modeling in both non-reacting and reacting turbulent flows \cite{Bode2021, Bode2023}. This approach has proven to be effective at accurately reconstructing fine-scale turbulent structures, a task where traditional supervised methods often fail. The adversarial training process significantly enhances the model generalization capabilities for out-of-distribution conditions, such as flows at higher Reynolds numbers \cite{Nista2024}. The framework versatility has since been demonstrated in high-ratio DNS data compression \cite{ZipGAN} and implementation at scale \cite{Nista2025}.
Additionally, physics-guided self-supervised models \cite{Dash2024} have shown encouraging results in capturing statistical properties such as dissipation spectra and structure functions, necessary for subgrid-scale modeling. More recently, transformers and DDPMs have emerged as powerful tools for capturing long-range dependencies in data \cite{Li2024, Shu2023}, though their high computational demands render them impractical for many canonical fluid dynamics problems.

Traditional machine learning models often struggle with unstructured mesh data in fluid simulations, as they typically assume uniform grids. This limitation hinders their ability to capture spatial relationships and generalize to complex geometries and meshes. GNNs extend traditional neural network architectures to operate on graphs, allowing them to process complex relational information. Graph convolutions (or message passing), which aggregate information from neighboring nodes to update node representations, form the foundation of many GNN models. Further enhancements like graph attention mechanisms \citep{GAT2018} assign learnable weights to prioritize important neighbors, while graph autoencoders \citep{Kipf2016} enable un/supervised learning of compact representations. GraphSAGE \citep{Hamilton2018} employs neighborhood sampling for efficient large-scale graph handling. PointNet \citep{Qi2017}, though not strictly a GNN, offers similar capabilities for processing unordered point cloud data, making it applicable to graph-based tasks. These techniques collectively provide a robust framework for handling the challenges posed by unstructured data in fluid simulations.

Battaglia et al. \citep{Battaglia2018} introduced a standardized formulation for graph neural networks (GNNs), outlining key design principles for such architectures. Since then, GNNs have been widely adopted to accelerate computational simulations across various domains, including molecular dynamics \cite{Li2022MD}, rigid body dynamics \cite{Bhattoo2022}, and continuum mechanics \cite{Pfaff2021}, utilizing both Eulerian and Lagrangian formulations. One of the pioneering efforts in this field is MeshGraphNet \citep{Pfaff2021}, a generalizable and scalable mesh-based model capable of predicting the dynamics of diverse physical systems. It employed an autoregressive encoder-processor-decoder architecture, effectively modeling flow and deformation in the Eulerian and Lagrangian domains, respectively. Building upon this, the multiscale GNN \citep{Lino2022} introduced a hierarchical approach, processing information across multiple scales through progressive graph downsampling and upsampling, akin to a U-Net architecture. This model significantly improved inference speed and enhanced stability for long rollout times, a critical factor in autoregressive simulations. 

GNNs have also been utilized for tasks such as interpretable compression and super-resolution of fluid simulation data on non-uniform grids. Barwey et al. \citep{Barwey2023} introduced a multiscale graph autoencoder that effectively captured the evolution of coherent structures within interpretable latent spaces. A leaner version of the model \citep{Barwey2025} featuring a synchronization layer was applied for learnable upsampling in snapshots obtained from Galerkin-based methods, analogous to $p$-refinement, where the polynomial degree of the basis functions is increased to improve resolution. Its effectiveness was demonstrated using two cases: Taylor-Green vortex and backward-facing step, and the geometry extrapolation capability of the model was also assessed. While GNN-based super-resolution has been demonstrated for non-reacting flows, extension to turbulent reacting flows on complex meshes introduces an additional challenge associated with the flame front. On coarse meshes, the reaction zone becomes artificially thickened, smearing out gradients in temperature and species and potentially distorting the coupling between turbulence and chemistry. Therefore, any super-resolution strategy intended for reacting flows must recover both small-scale flow structures and the flame structure. The objective of this study is to develop a GNN-based framework for reconstructing unresolved features of turbulent reacting flows on complex meshes, targeting the full set of thermochemical and hydrodynamic scalar fields, including accurate species and temperature profiles across the flame front. The key contributions of this work are as follows:

\begin{itemize}
    \item We develop a GNN-based framework with message-passing layers for interpolation-free super-resolution of three-dimensional turbulent reacting flow data on non-uniform structured meshes.
    \item For unstructured meshes, we formulate super-resolution in a manner analogous to \(p\)-refinement in spectral element methods: the coarse-mesh solution is represented on a higher-order mesh, and the GNN infers unresolved content directly at the target high-resolution degrees of freedom.
    \item We assess the framework on two applications: a reacting channel flow on a non-uniform structured mesh and a practically relevant lean hydrogen internal combustion engine configuration on an unstructured mesh.
\end{itemize}

The remainder of this paper is organized as follows. Section~\ref{sec:dldetails} presents the graph generation and deep learning framework. The datasets, featuring a structured non-uniform mesh and an unstructured mesh are described in Section~\ref{sec:datasets}. The results are presented and discussed in Section~\ref{sec:results}, and the key findings as well as future directions are summarized in Section~\ref{sec:conclusions}.

\section{Graph generation and deep learning methodology\label{sec:dldetails}} 

\begin{figure*}[h!]
\setlength{\unitlength}{1cm}
\vspace{-1.3 cm}
\includegraphics[trim={0cm 0.5cm 0cm 1.3cm},clip,width=15cm]{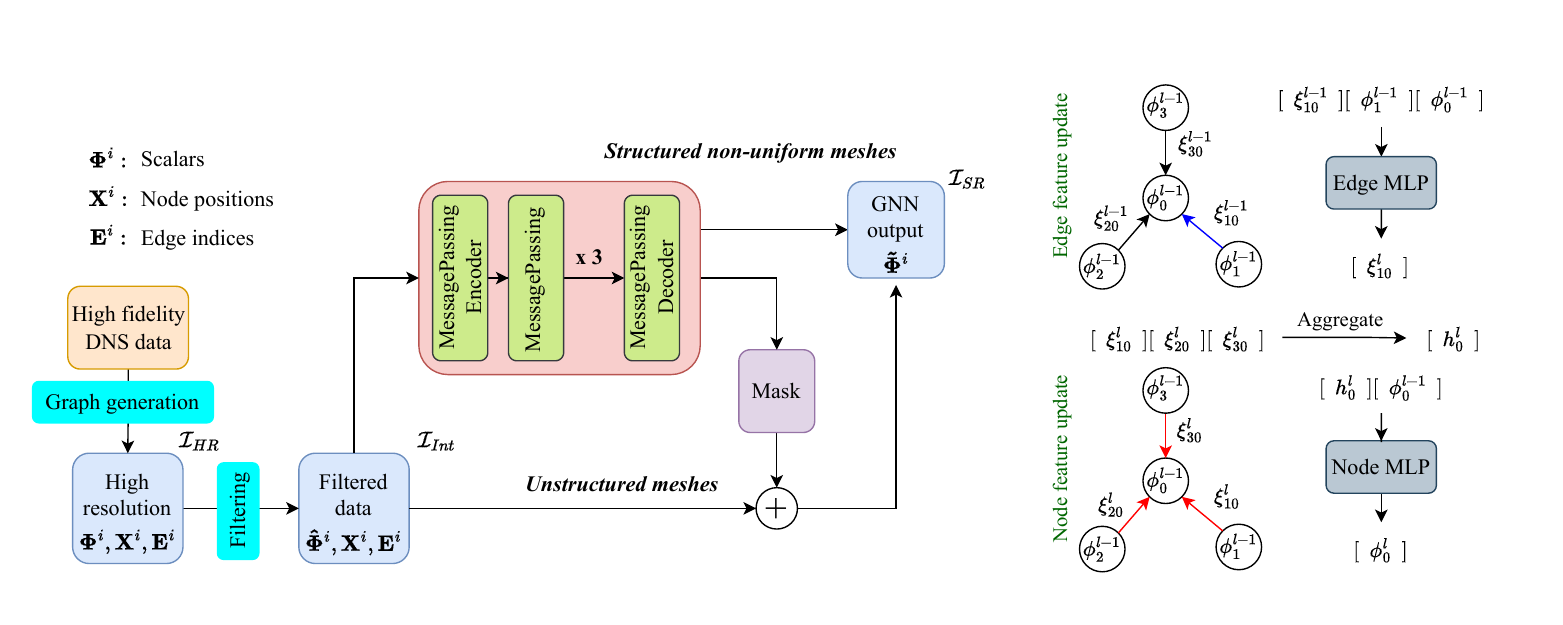}
\caption{\footnotesize A schematic of the GNN framework for super-resolution {\bf(left)}, and message passing layer {\bf(right)} showing edge feature update, edge feature aggregation, and node feature update.}
\label{fig:schematic}
\end{figure*}

A deep learning-based super-resolution framework takes in a coarse field \( \mathcal{I}_{LR} \) and predicts a corresponding high-resolution reconstruction \( \mathcal{I}_{SR} \). During training, paired data are constructed by downsampling or filtering a high-resolution reference \( \mathcal{I}_{HR} \) to obtain \( \mathcal{I}_{LR} \). The network is then optimized by minimizing a loss function defined on the discrepancy between \( \mathcal{I}_{SR} \) and \( \mathcal{I}_{HR} \). During inference, only \( \mathcal{I}_{LR} \) is provided, and the trained model is used to generate \( \mathcal{I}_{SR} \) without access to \( \mathcal{I}_{HR} \). In practice, 3D fields are often partitioned into smaller subdomains to cope with memory constraints during both training and inference.

These subdomains are compatible with convolutional neural networks (CNNs) when structured uniform meshes are used, which do not require the encoding of spatial information about the nodes. For complex meshes, data from each subdomain must be reformulated from structured four-dimensional tensor data into a graph representation, comprising nodes, edges, and corresponding features, to be processed by graph neural networks (GNNs). Next, we outline the dataset generation process for non-uniform and unstructured mesh data, along with details of the network architecture and training setup. A schematic of the proposed framework is shown in Fig. \ref{fig:schematic}.

\subsection{Structured non-uniform mesh data\label{sec:nudata}} 

For structured non-uniform mesh data, $\mathcal{I}_{HR} \in \mathbb{R}^{N_{s} \times N_x \times N_y \times N_z}$, where $N_s, N_x, N_y, N_z$ denote the number of variables and nodes/cells along each spatial direction. 
In this work, $\mathcal{I}_{HR}$ is filtered using average pooling by 
taking the arithmetic mean of the variables in each pooling region
to produce lower-resolution data ${\mathcal{I}}_{LR}$. Pairs of $\mathcal{I}_{HR}$ and ${\mathcal{I}}_{LR}$ are then used to train the super-resolution model. ${\mathcal{I}}_{LR}$ can be upsampled back to the original resolution using standard interpolation schemes, and the result is denoted as $\mathcal{I}_{Int}$. Here, we have used trilinear interpolation for structured non-uniform meshes.
Even though $\mathcal{I}_{Int}$ has the same resolution as $\mathcal{I}_{HR}$, small scale structures that were filtered out during the downsampling process are not recovered. $\mathcal{I}_{Int}$ can be used as an input in the pre-upsampling mode of super-resolution, following the methodology in \citep{Bode2021}. 

Due to the memory limitations imposed by the employed GPU, subdomains of size $N_s \times 32^3$ are extracted from both $\mathcal{I}_{HR}$ and $\mathcal{I}_{Int}$. These subdomains are flattened to ensure compatibility with the \texttt{PyTorch-Geometric} framework \cite{Fey2025}, and are denoted as ${\bf \Phi}^i$ and ${\bf \hat{\Phi}}^i$, respectively, where each subdomain has a dimensionality of $N_n \times N_s$, with $N_n~\text{and}~i$ respectively denoting the number of nodes in a subdomain and the subdomain index. Each subdomain ${\bs \Phi}^i$ is further decomposed as 
\begin{equation}
{\bs \Phi}^i = \{{\bs \phi}^i_1, {\bs \phi}^i_2, \dots, {\bs \phi}^i_{N_n}\}^T
\end{equation}
where at each node 
\begin{equation}
{\bs \phi}^i_j = \{\bs u^i_{j}, T^i_j, p^i_j, \rho^i_j, \bs Y^i_{j}\}
\end{equation} 
represents the normalized velocity vector, temperature, pressure, density, and species mass fractions. An analogous notation is used for the ${\bf \hat{\Phi}}^i$ data. Additionally, positional data of the nodes in subdomain $i$ are captured as 
\begin{equation}
{\bs X}^i = \{{\bs x}^i_1, {\bs x}^i_2, \dots, {\bs x}^i_{N_n}\}^T,
\end{equation}
which remains consistent across both high-resolution and interpolation. A neighborhood is defined for each node by considering its one-hop neighbors along the primary axes, resulting in a maximum of 7 neighbors per node, including the self-link. Depending on the node position within the subdomain, the number of neighbors can vary between 4 (corner nodes) and 7 (internal nodes). The set of neighbors of a node $j$ in subdomain $i$ is denoted as $\mathcal{N}^i(j)$, with the adjacency of the graph represented by the ordered pairs 
\begin{equation}
{\bs E}^i = \{(j, k)~|\{j,k\} \in {\bf \Phi}^i,~k \in {\mathcal{N}}^i(j)\}.
\end{equation}
Each subdomain is thus represented as a graph $\mathcal{G}^i = (V^i, {\bs E^i})$, where the nodes $V^i = ({\bf \hat{\Phi}}^i, {\bf \Phi}^i)$ correspond to the input and output features used for training. In line with the \texttt{PyTorch-Geometric} framework, the directed edge features are initialized as 
\begin{equation}
{\bs \xi}^i_{jk} = \{{\bs \phi}^i_j - {\bs \phi}^i_k, \,\, {\bs x}^i_j - {\bs x}^i_k, \,\, ||{\bs x}^i_j - {\bs x}^i_k||^2_2\},
\label{eq:edge-feat-gen}
\end{equation}
encapsulating the differences in scalar quantities, relative positions, and the Euclidean distance between connected nodes.

\subsection{Unstructured mesh data\label{sec:usdata}}

Graph dataset generation for unstructured mesh data requires an explicit definition of the neighborhood mapping between elements. In this work, we focus on a dataset obtained from a spectral element solver (details provided in the next section), where the 3D domain is discretized using conforming hexahedral elements, within each of which eight Gauss-Legendre-Lobatto quadrature points per direction were used for the chosen $p=7$ order Lagrange polynomial approximation. Each spectral element is treated as a graph, with the element nodes serving as graph nodes and their `intra-element' neighbors defining the local connectivity. 

The low-resolution representation of an element is obtained by selecting only the corner nodes, i.e. two nodes per direction, effectively downsampling the scalars from 7th-order (\( p = 7 \)) to the first-order (\( p = 1 \)) elements. The values are then interpolated back to \( p = 7 \) space using inverse distance weighting. With eight nodes per direction, each spectral element has 512 nodes. For training, batches of {$N_b~=~768$} elements sampled across the domain are concatenated to form a graph. With $N_s=14$ variables per node, \( {\bf \Phi}^i \) and \( {\bf X}^i \) in this case have dimensions of \( (N_b\times512) \times 14 \) and \( (N_b\times512) \times 3 \), respectively. It is important to emphasize that the upsampling ratio in the current work for unstructured meshes is approximately 8$\times$, which is arduous for turbulent reacting flows. 
Although the data extraction strategy has been developed in the context of the spectral element method, it can be extended to other unstructured discretizations, such as finite volume methods, wherein graph nodes may represent cell centers and edges may be defined through face adjacency or stencil-based connectivity.

\subsection{Graph neural network (GNN) details \label{sec:gnndetails}} 
The employed model\footnote{A simplified version of the GNN will be made available upon acceptance.} is based on {an encoder-processor-decoder architecture, as described in Pfaff et al.~\citep{Pfaff2021}}. In the paradigm of graph-based deep learning models, every layer can be formulated as a message-passing (MP) layer. For this work, we use a specialized version of MP tailored to our datasets. The model comprises {five} MP layers, where the first layer acts as the encoder and the last one as the decoder. In each layer, the edge features (${\bs \xi}^{i,l-1}_{jk}$) from the previous layer are first passed through the $l^\text{th}$ layer of a multi-layer perceptron ($\text{MLP}^l_e$), yielding updated edge features ${\bs \xi}^{i,l}_{jk}$. For the initial layer ($l = 0$), the edge features are initialized as described in Equation (\ref{eq:edge-feat-gen}). The edge-feature update process is governed by 
\begin{equation} 
{\bs \xi}_{jk}^{i,l} = \text{MLP}_e^l ({\bs \xi}_{jk}^{i,l-1}, {\bs \phi}_j^{i,l-1}, {\bs \phi}_k^{i,l-1}) .
\label{eq:mp_edge_update}
\end{equation}
Subsequently, the edge features corresponding to all edges connected to a node are aggregated using a permutation-invariant function. Here, we employed the mean as the aggregation function, 
\begin{equation} 
{\bs h}_{j}^{i,l} = \frac{1}{{|\cal N}(j)|}\sum_{k \in {\cal N}(j)} {\bs \xi}_{jk}^{i,l}.
\end{equation}  
The aggregated edge features are then concatenated with the node features and passed through the $l^\text{th}$ layer of the node MLP, denoted by $\text{MLP}_{\bs \phi}^l$, to generate the updated node features, 
\begin{equation} 
{\bs \phi}_{j}^{i,l} = \text{MLP}_{\bs \phi}^{l} ({\bs h}_{j}^{i,l},{\bs \phi}_j^{i,l-1}).
\end{equation} 
The dimensionality of the inputs and outputs for the MLPs is determined by the dimensions of the edge and node features, with the number of intermediate features fixed at {128} across all MLPs, as demonstrated in \cite{Pfaff2021, Barwey2025}. The exponential linear unit (ELU) activation function is chosen because of its nonlinearity across real space and near-zero mean activation, facilitating the efficient learning of complex relationships in scientific datasets.
 
Similar to data generation, the application of the network to interpolated data to generate the super-resolved output (\( {\bf \tilde{\Phi}}^i \)) differs between the two types of meshes considered. For non-uniform structured meshes, the subdomains do not overlap, ensuring that each nodal location belongs to a single subdomain. Consequently, {with the model parameters ($\theta$)}, the upsampling process can be expressed as  
\begin{equation}  
{\bf \tilde{\Phi}}^i = \text{GNN}({\bf \hat{\Phi}}^i,{\bs X}^i,{\bs E}^i;\theta), 
\label{eq:case1_sr}  
\end{equation}   
which ideally facilitates information propagation across larger scales, up to five-hop neighbors for most nodes. 

On the other hand, the GNNs for the unstructured mesh data are processed element-wise \cite{Barwey2025}. Within an element, there are two types of nodes: those belonging to both $p=1$ (\glr) and $p=7$ (\ghr,~ \gint~, and~\gsr) and the ones present only in $p=7$ elements. Here, the super-resolution model should only predict values at nodes that are not part of \glr. This is incorporated by introducing a masking variable \( {\bs M}^i \) which modifies the upsampling equation as follows  
\begin{equation}  
{\bf \tilde{\Phi}}^i = {\bf \hat{\Phi}}^i + {{\bs M}^i \, \odot\,\,} \text{GNN}({\bf \hat{\Phi}}^i,{\bs X}^i,{\bs E}^i;\theta),
\label{eq:case2_sr}  
\end{equation}   
Here, $\odot$ represents element-wise multiplication, and \( {\bs M}^i \) is set to zero for the first type of nodes (located at the element vertices) and to unity elsewhere. 

\subsection{Convolutional neural network (CNN) baseline \label{sec:traindetails}}
To establish a meaningful reference for evaluating the performance of the proposed GNN-based framework, we also include a baseline CNN for the structured mesh case. This choice was deliberate, as convolution and patch-attention models inherently assume structured, uniformly spaced meshes, which makes them ill-suited for the irregular grids encountered in many practically-relevant simulations. To ensure a fair comparison, the CNN is designed to match the GNN in complexity, with an identical depth of five layers and 128 intermediate channels, thereby isolating the effect of the network architecture rather than the model capacity. The workflow for this baseline involves several steps: data from the non-uniform mesh is first mapped onto a uniform mesh with the same number of grid points, the CNN is trained and evaluated entirely in this uniform space, and the predicted outputs are then mapped back to the original resolution for post-processing and analysis. During training it was observed that the validation loss in this case stagnated at 750 epochs in almost all instances of the hyperparameter sweep. 

\subsection{Training \label{sec:traindetails}}

The training process is implemented using the distributed data-parallel framework provided by PyTorch \cite{Li2020DDP}, enabling compatibility with multi-node multi-GPU architectures. The loss function is formulated as the mean squared error (MSE) of the scalar fields, which is computed for each of the $N_n$ subdomains as 
 \begin{equation}   
 \mathcal{L}^i = \frac{1}{N_n}||{\bf \hat{\Phi}}^i - {\bf {\Phi}}^i ||^2_2.   
 \label{eq:lossfn}   
 \end{equation}   
The model was trained on approximately 90\% of the subdomains, with the remaining 10\% reserved for validation to monitor generalization and prevent overfitting. The initial learning rate was set to $6\times 10^{-4}$ and manually reduced by half whenever the loss stagnated over a window of 100 epochs. {Training and inference were conducted on compute systems of the  J\"{u}lich Supercomputing Centre featuring nodes with either four NVIDIA GH200 superchips or four NVIDIA A100 GPUs. The GH200 superchips contain Hopper GPUs with 96\,GB HBM3 memory each. The A100 GPUs provide 40\,GB HBM2 memory each.} The datasets details are presented in the next section.

\section{Datasets\label{sec:datasets}}
\begin{table}[h]
\caption{\label{tab:diff}{Table outlining differences between Cases 1 and 2.}}
\centering
\fontsize{8pt}{9pt}\selectfont
\begin{tabular*}{\columnwidth}{@{\extracolsep{\fill}}|p{1.5cm}|p{1.9cm}|p{1.9cm}|} 
\hline
         & \textbf{Case 1}                 & \textbf{Case 2}       \\ \hline\hline
Mesh     & Structured non-uniform & Unstructured \\ \hline
Geometry & Canonical              & Real-scale   \\ \hline
Solver   & Compressible           & Low Mach     \\ \hline
Nodes/instance    & 63.06 million               & 3.78 billion     \\ \hline
DoFs/node & 29                    & 14           \\ \hline
Demonstration & Comparison with CNN baseline & OOD extrapolation.\\
\hline
\end{tabular*}
\end{table}

\subsection{Case 1: Reacting channel flow}
\begin{figure}[h!]
\centering
\setlength{\unitlength}{1cm}
\includegraphics[width=7.2cm]{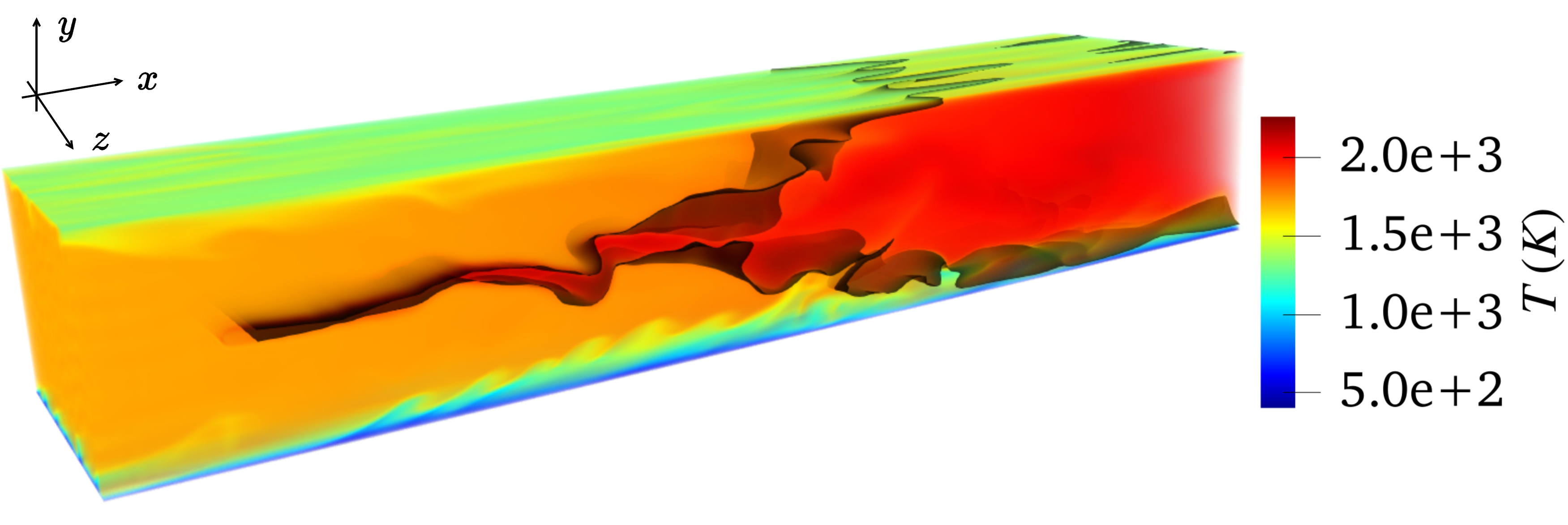}

\caption{\footnotesize Volumetric rendering of temperature inside the channel, with the flame surface represented in black color. }
\label{fig:channel-vol}
\end{figure}

The first dataset \cite{BlastNet, Chung2022} includes twelve time instants of the fully-compressible reactive flow generated using direct numerical simulation (DNS) with the high-order finite difference code NTMIX-CHEMKIN \cite{Baum1994}. The objective of the study \citep{Jiang2021} was to investigate the flame-wall interactions of premixed methane/air flames diluted by hot combustion products. The computational domain features turbulent inflow with coupled velocity and temperature fluctuations, non-reflecting outflow and periodicity along the $z$-direction. The temperature of the isothermal walls varies smoothly from the inlet mixture temperature to \SI{400}{K} and \SI{1200}{K} on the lower and upper wall, respectively. The domain has an aspect ratio of 12:2:3 with a resolution of $1001\times251\times251$. The grid spacing is uniform along $x$ and $z$, and refined near the walls in $y$ to capture wall effects. Chemical kinetics is described by a 23-species 205-reactions analytically-reduced mechanism obtained from GRI 3.0. A cylindrical hot patch on the mid plane close to the inflow stabilizes a 3D turbulent V-flame inside the channel. {Each instance within the dataset comprises 29 fields {(velocity, temperature, pressure, density, and mass fractions)}, where each scalar value at a node is a degree of freedom (DoF). {Figure \ref{fig:channel-vol} provides a visual overview of the instantaneous flow field for the reacting channel.} The model was trained for 5900 epochs, with each epoch requiring on average \SI{13.3}{s} using 16 superchips (with an average load of 0.411 billion DoFs/superchip/epoch including I/O).}

\subsection{Case 2: Lean hydrogen internal combustion engine}

\begin{figure}[h]
\centering
\setlength{\unitlength}{1cm}
\includegraphics[width=7.0cm]{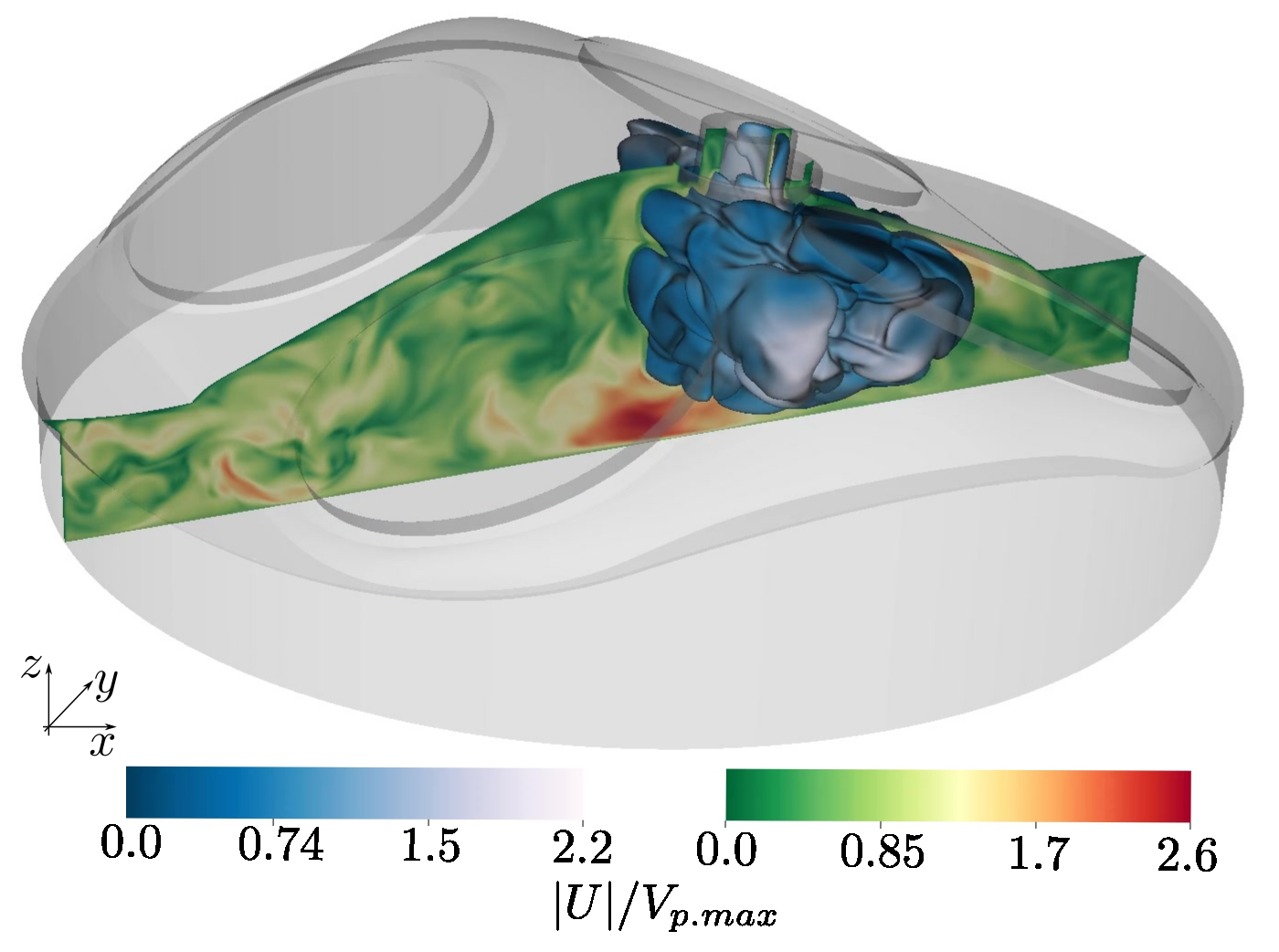}

\caption{\footnotesize Flame front isosurface superimposed on the normalized velocity magnitude distribution on the engine midplane.} 
\label{fig:engine-vol}
\end{figure}

For the unstructured mesh case, we consider a dataset obtained from a state-of-the-art DNS of lean hydrogen combustion in the TU Darmstadt optical engine \citep{Danciu2025}. The computational domain is a single-cylinder, spark-ignition engine with a pent-roof, four-valve head. The engine operates at \SI{800}{rpm} with an intake pressure of \SI{0.4}{bar} and an equivalence ratio of $\phi=0.4237$ \cite{Welch2024}. Kinetics is described by the detailed mechanism of \citep{Burke2012} comprising 9 species in 23 reactions. The DNS is performed using nekCRF \citep{Kerkemeier2024}, a spectral element low Mach number reacting flow solver based on nekRS \cite{nekrs} that is optimized for massively parallel heterogeneous architectures. {Figure \ref{fig:engine-vol} shows a representative snapshot of the complex flow environment within the engine geometry.}

The model is trained on a state taken at 8.8 crank angle degrees after the start of ignition when the ignition kernel has formed a propagating front and the domain contains a significant fraction of burnt and unburnt regions. The computational domain was discretized using 7.38 million hexahedral spectral elements and the solution is approximated with 7-th order polynomials. With 14 scalars per node, the pipeline is expected to handle approximately 52.92 billion DoFs, presenting a substantial challenge for scientific machine learning. {The model was trained for 2750 epochs on this dataset. Utilizing 64 superchips with an average load of 0.744 billion DOFs/superchip/epoch including I/O, the average wall time per epoch was \SI{58.6}{s}.} {The differences between the two cases are summarized in Table \ref{tab:diff}.}

\section{Results\label{sec:results}}

\begin{sidewaysfigure*}[p]
    \setlength{\unitlength}{1cm}
\centering
\includegraphics[trim={0cm 0cm 0cm 0cm},clip,width=20cm]{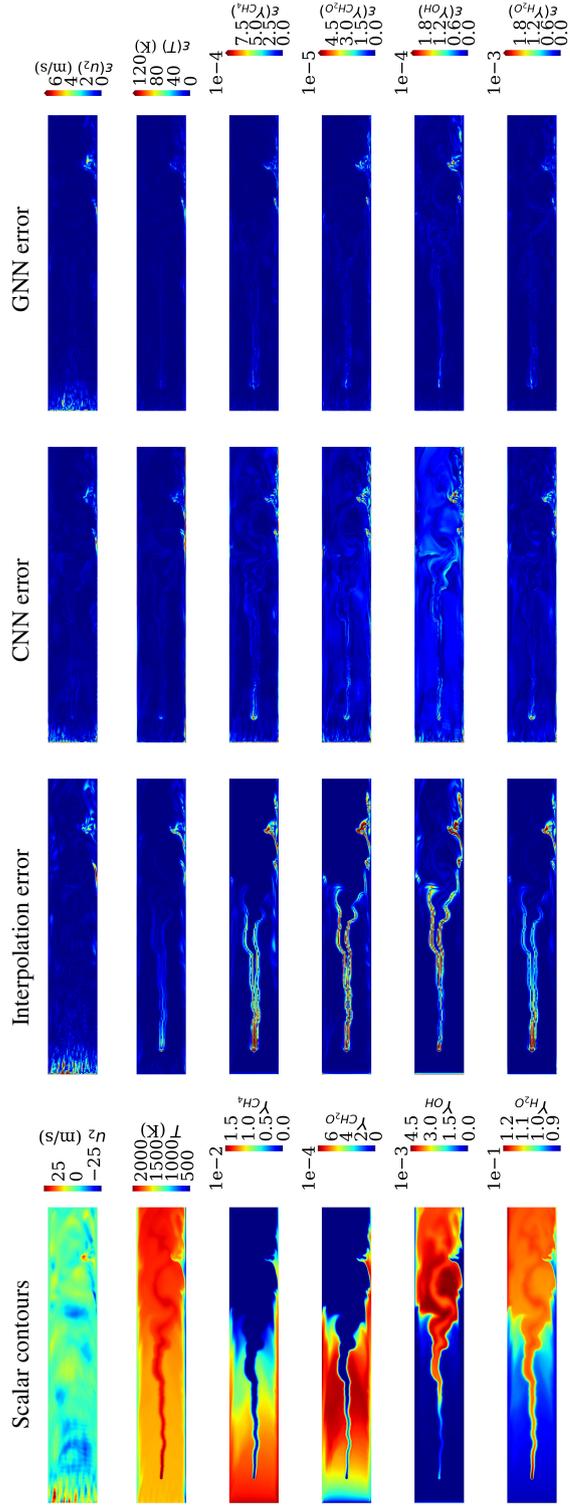}
\put(-18.8,7.2){{Scalar contours}}
\put(-13.5,7.2){{Interpolation error}}
\put(-8.7,7.2){{CNN error}}
\put(-4.1,7.2){{GNN error}}

\caption{\footnotesize Error reduction by GNN compared to trilinear interpolation and CNN baseline for Case 1. {\bf Column 1:} Contours of key variables shown for the center-plane along $z$. {\bf Columns 2, 3, and 4:} Contours of absolute errors $\epsilon$ between ground truth (DNS solution) and the results of interpolation, CNN, and GNN.}
\label{fig:err-contours}
\end{sidewaysfigure*}

For the analysis of the GNN super-resolution (SR) framework in Case 1, we consider a checkpoint that was not used in the training process. The inference procedure differs from training in several key respects. First, in addition to the $32^3$ subdomains described earlier, we augment the domain with a one-cell-thick buffer region in every direction. During training, each node in the $32^3$ subdomain can have up to seven neighbors (two along each spatial direction plus a self-connection). However, depending on its position within the subdomain, some nodes will have only four, five, or six neighbors, which can introduce windowing artifacts during inference. To mitigate this effect, we pad the full simulation domain ($1001\times 251\times 251$) with one layer of nodes in every direction, expanding it to $1003\times 253\times 253$. This allows us to use $34^3$ subdomains during inference and retain only the internal nodes, ensuring that every retained node consistently has seven neighbors. Furthermore, appropriate padding is applied to maintain consistency in boundary conditions. At the full checkpoint level, periodic padding is implemented along the $z$-direction, while replication padding is applied along the $x$- and $y$-directions to preserve the integrity of the computational domain.

\begin{figure*}[!h]
\centering
\setlength{\unitlength}{1cm}
\includegraphics[trim={0cm 0.7cm 0cm 0cm},clip,width=14cm]{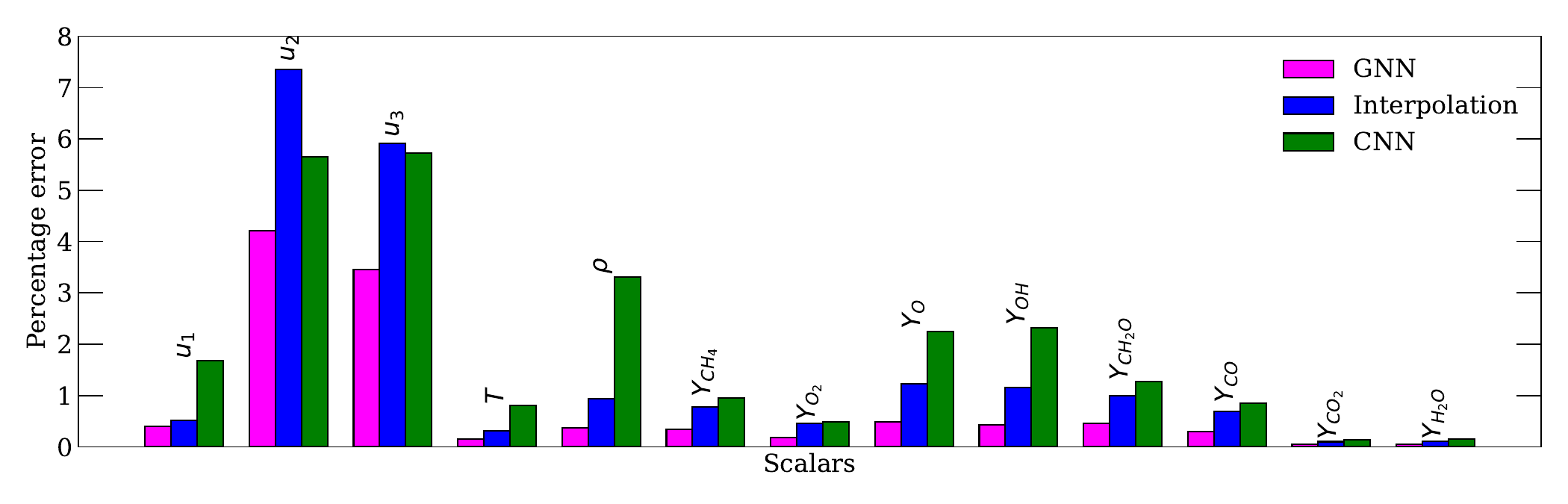}
\caption{\footnotesize{Cumulative error percentage for selected scalars across all upsampling methods in Case 1.}}
\label{fig:channel-global-err}
\end{figure*}

The first column in Fig. \ref{fig:err-contours} presents the contours of key scalar fields extracted on the center plane along the $z$-direction. These include the transverse velocity component ($u_2$), temperature, and the mass fractions of \ce{CH4}, \ce{CH2O}, \ce{OH}, and \ce{H2O}. The other columns compare absolute errors {computed} between the interpolated field (\gint) and the high-resolution ground truth (\ghr) (second column), and with respect to the CNN-based output (\gcnn) and the GNN-based output (\gsr) (third and fourth columns, respectively). For $u_2$, the scattered discrepancies seen in the interpolated field are reduced in the CNN output and are further suppressed in the GNN reconstruction. A similar trend is observed for temperature, where the localized errors near the flame front and bottom wall are visibly diminished in \gsr. In comparison, the CNN reduces peak errors at the flame but introduces spatially smeared discrepancies across the domain. For scalars present across different zones of the premixed flame, from reactants to products, the GNN consistently minimizes differences near the flame and bottom wall wherever they occur. The CNN output exhibits similar behavior near the flame but retains broader errors throughout the domain. The erroneous region near the aft end of the domain along the bottom wall, prominent in \gint, is moderately reduced in the CNN and substantially attenuated in the GNN. This demonstrates that upsampling using the GNN results in a better match with the ground truth than interpolation and the CNN (intended to be a data-driven baseline), particularly in regions critical for extracting fundamental insights from DNS data.

\setlength{\tabcolsep}{3pt}

\newcolumntype{B}{>{\bfseries}c}
\newcommand{\gnncolor}{\rowcolor{magenta!25}}
\newcommand{\intcolor}{\rowcolor{blue!25}}
\newcommand{\cnncolor}{\rowcolor{green!25}}

\begin{table*}
\caption{\label{tab:region-errors} Percentage error across regions for each scalar across different upsampling approaches.}
\label{tab:channel-local-err}
\centering
\begin{tabular}{|B|B|c|c|c|c|c|c|c|c|c|c|c|c|c|}
\hline
Region & Model & $u_1$ & $u_2$ & $u_3$ & $T$ & $\rho$ & $Y_{CH_4}$ & $Y_{O_2}$ & $Y_{O}$ & $Y_{OH}$ & $Y_{CH_2O}$ & $Y_{CO}$ & $Y_{CO_2}$ & $Y_{H_2O}$ \\
\hline
\multirow{3}{*}{Bottom wall} & GNN & 1.17 & 9.02 & 5.22 & 0.58 & 0.93 & 0.47 & 0.25 & 2.90 & 2.80 & 0.77 & 0.72 & 0.07 & 0.11 \\
& Interpolation & 1.76 & 13.85 & 8.39 & 1.17 & 2.59 & 0.78 & 0.41 & 4.34 & 4.72 & 1.07 & 1.17 & 0.08 & 0.14 \\
& CNN & 6.22 & 14.67 & 11.44 & 3.44 & 8.22 & 1.31 & 0.70 & 18.82 & 16.70 & 1.98 & 2.20 & 0.28 & 0.31 \\
\hline
\multirow{3}{*}{Center} & GNN & 0.23 & 3.01 & 2.90 & 0.07 & 0.12 & 0.29 & 0.17 & 0.36 & 0.30 & 0.35 & 0.22 & 0.05 & 0.04 \\
& Interpolation & 0.22 & 5.78 & 5.20 & 0.15 & 0.19 & 1.02 & 0.59 & 1.17 & 1.04 & 1.23 & 0.69 & 0.13 & 0.11 \\
& CNN & 0.51 & 3.07 & 3.08 & 0.13 & 0.79 & 0.73 & 0.39 & 1.42 & 1.49 & 0.98 & 0.51 & 0.08 & 0.10 \\
\hline
\multirow{3}{*}{Top wall} & GNN & 0.54 & 5.38 & 3.21 & 0.11 & 0.15 & 0.25 & 0.12 & 0.57 & 0.55 & 0.33 & 0.25 & 0.03 & 0.04 \\
& Interpolation & 0.85 & 8.59 & 5.28 & 0.19 & 0.24 & 0.27 & 0.14 & 0.73 & 0.88 & 0.38 & 0.31 & 0.03 & 0.04 \\
& CNN & 3.45 & 10.27 & 8.28 & 1.03 & 1.95 & 0.91 & 0.48 & 2.44 & 3.35 & 1.02 & 0.95 & 0.16 & 0.13 \\
\hline

\end{tabular}
\end{table*}

To assess the overall accuracy of the trained models, we compute the domain-integrated error for each scalar field normalized by the corresponding ground truth magnitude, expressed as  
\begin{equation}
    \text{Cumulative error} = 100 \times \frac{\sum|\mathcal{I}_{rec.} - \mathcal{I}_{HR}|}{\sum|\mathcal{I}_{HR}|},
    \label{eqn:cumulative-err}
\end{equation}  
where $\mathcal{I}_{rec.}$ and $\mathcal{I}_{HR}$ denote the reconstructed and high-resolution fields, respectively, and the summation is performed over the entire computational domain. As shown in Fig. \ref{fig:channel-global-err}, the GNN significantly reduces the error introduced by interpolation, with reductions ranging from 20\% to 40\% for hydrodynamic scalars and even larger improvements observed for thermochemical scalars. In contrast, the CNN baseline introduces additional error {for all scalars barring transverse and spanwise velocity components}, performing worse than interpolation overall.

To investigate the spatial distribution of the reconstruction errors, we partition the domain into three regions according to the wall-normal location: near-wall zones adjacent to the bottom and top boundaries (first and last 10\%) and a central region covering the intermediate 80\%, as summarized in Table \ref{tab:channel-local-err}. In the near-wall regions, the GNN consistently yields the lowest errors, followed by interpolation, with the CNN performing worst. In the central region, the CNN produces lower errors than interpolation for most scalars, which is consistent with the fact that the convolution filters/kernels are naturally aligned with the locally uniform mesh and can better exploit the smooth spatial correlations there than interpolation. Nevertheless, the errors incurred by the GNN remain lower than those of both interpolation and the CNN, highlighting the robustness of the GNN model in accurately reconstructing diverse scalar fields across the entire domain and reinforcing its superiority over the CNN baseline. Consequently, the following discussion focuses solely on comparisons between the GNN output and interpolation.

\begin{figure}[h!]
\centering
\setlength{\unitlength}{1cm}
\hspace{-0.5cm}
\includegraphics[width=7.2cm]{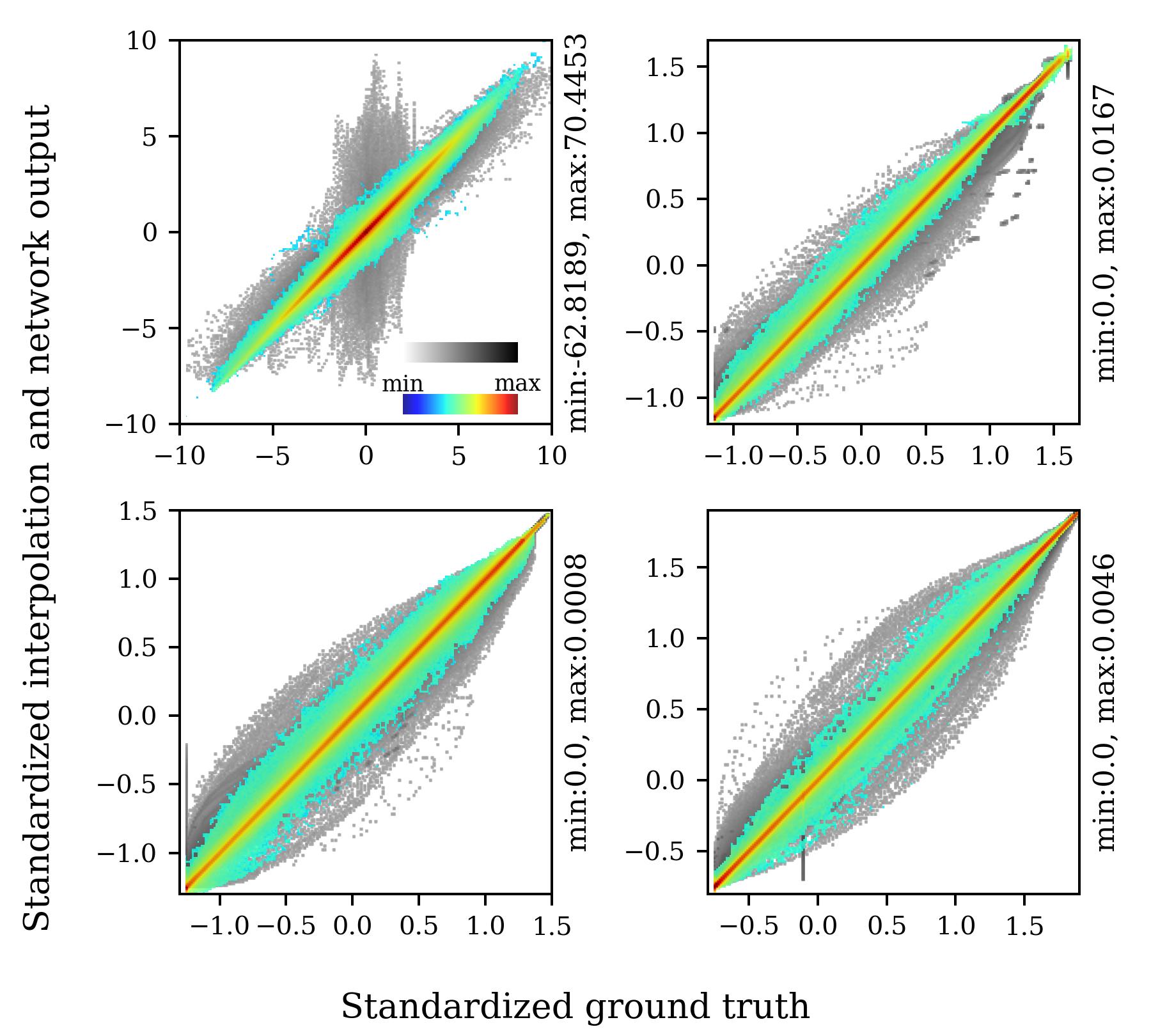}
\put(-5.8,5.9){{$u_2$}}
\put(-2.5,5.9){{$Y_{\ce{CH4}}$}}
\put(-5.8,2.8){{$Y_{\ce{CH2O}}$}}
\put(-2.5,2.8){{$Y_{\ce{OH}}$}}
\caption{\footnotesize Joint PDFs computed between interpolation/network output and ground truth for transverse component of velocity and mass fraction of three species for Case 1. Grayscale color scheme is used for interpolation and jet colormap for the GNN output.
}
\label{fig:jpdfs}
\end{figure}
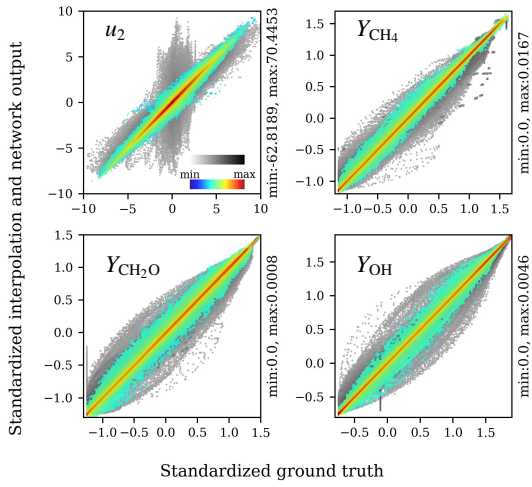

To further evaluate the framework, we compare the joint probability density functions (PDFs) of \gsr~ and \gint~ with \ghr ~(Fig. \ref{fig:jpdfs}). This analysis goes beyond point-wise error metrics and offers a more comprehensive view of how well each approach captures the underlying data distribution. For $u_2$, \gint~ exhibits significant deviations, even at the mean values, which are noticeably streamlined by the trained GNN, although minor discrepancies remain. Similarly, for the mass fractions of \ce{CH4}, \ce{CH2O}, and \ce{OH}, the differences observed in \gint~ are substantially reduced in \gsr. The model effectively captures the complex statistical dependencies within the data, achieving reasonable accuracy on the non-uniform mesh.

\begin{figure}[h!]
\centering
\hspace{-0.7cm}
\includegraphics[trim={0cm 0.2cm 0cm 0.2cm},clip,width=7.2cm]{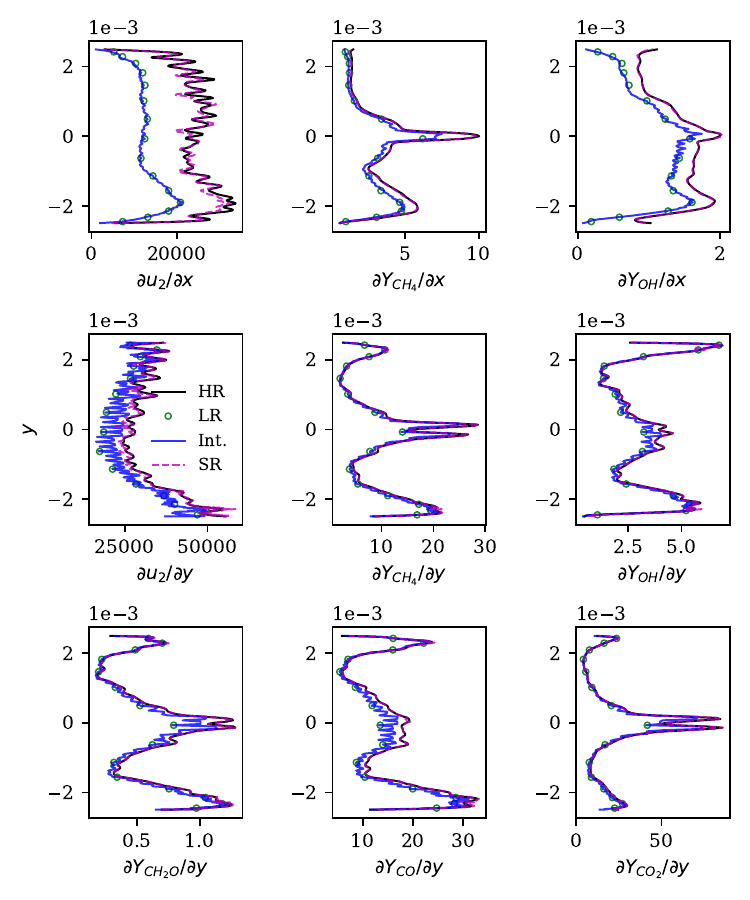}
\caption{\footnotesize Comparison of the root mean squared profiles of scalar gradients plotted along the transverse direction ($y$). {\bf Row 1:} gradients in the streamwise direction ($x$) and {\bf Rows 2,3:} gradients in the transverse direction ($y$).}
\label{fig:grad-rms}
\end{figure}

Reliable gradient reconstruction is essential for super-resolution in turbulent reacting flows since accurate gradients are critical to maintaining physical consistency, especially in regions with sharp variations like flame fronts. High-frequency details, which are often lost in coarse representations, depend on accurate gradient estimations to ensure realistic high-resolution predictions. Additionally, improved gradient reconstruction strengthens large eddy simulations (LES) by enhancing subgrid-scale (SGS) models, leading to better turbulence closure and more accurate flow simulations. To evaluate this, we plot the root mean squared (RMS) profiles of the gradients of selected scalars throughout the domain in Fig. \ref{fig:grad-rms}, focusing on both the streamwise and transverse directions.

In the first row, representing gradients along the streamwise direction, significant discrepancies between \gint~ and \gsr~ are observed. The trained GNN effectively reduces these errors. \gsr~ aligns closely with \ghr, except for minor deviations in $\partial{u_2}/\partial{x}$. The second and third rows focus on the gradients along the transverse direction, where the grid is stretched. Interpolation introduces noticeable fluctuations in these gradients owing to non-uniform spacing. However, the network effectively mitigates these artifacts, reducing the discrepancies between the interpolated data and the ground truth. This showcases its ability to recover accurate gradients despite grid-induced distortions.

\begin{figure*}[h!]
\setlength{\unitlength}{1cm}
\includegraphics[trim={0cm 0cm 0cm 0cm},clip,width=14.75cm]{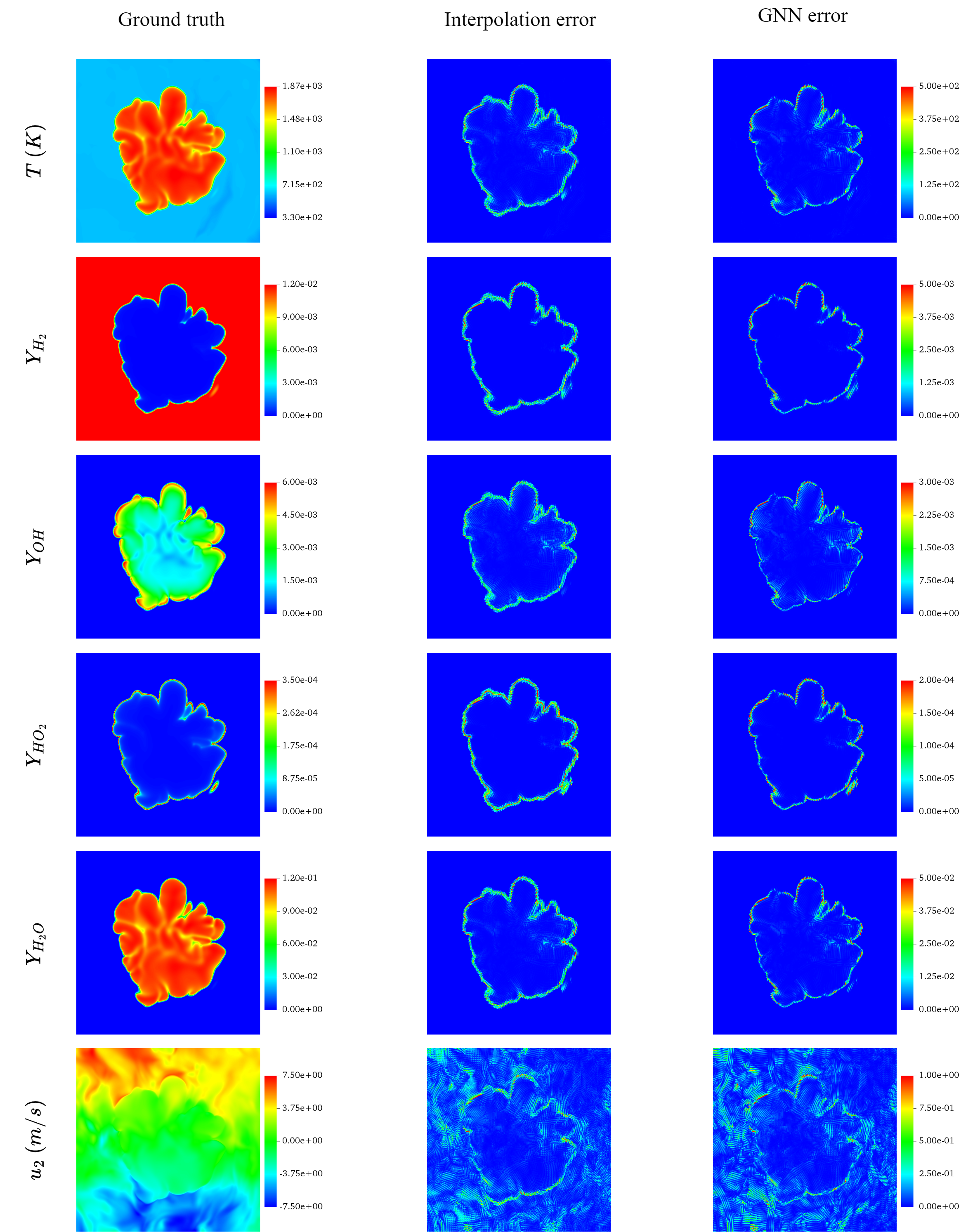}

\caption{\footnotesize Error reduction by GNN compared to interpolation for Case 2. {\bf Column 1:} Contours of key scalars on a slice approximately dividing the flame front into two equal halves along the $z$ direction. Deviation from ground truth is presented as absolute error for interpolation ({\bf Column 2}) and GNN output ({\bf Column 3}) computed for temperature, mass fractions of \ce{H2}, \ce{OH}, \ce{HO2}, and \ce{H2O}, and $y-$velocity.}
\label{fig:eng-err}
\end{figure*}

We shift our focus to Case 2 featuring a geometry of practical relevance simulated on an unstructured mesh. Unlike Case 1, padding is omitted here to concentrate on achieving super-resolution in an intra-element sense. It is worth noting that the internal combustion engine setup does not reach a statistically stationary state, and the system behavior varies significantly in time owing to the compression by the piston and the advancing flame front. Therefore, training and testing on snapshots at different CAD values inherently serves as a test of generalizability. Once inference is performed by treating each element as a separate graph, the resulting fields are stitched together using a direct stiffness summation, in which the predicted contributions at nodes common to multiple elements are gathered and combined into a single value. This assembly step enforces a unique, consistent value for each scalar at each shared node and yields a globally continuous reconstructed field across the entire mesh. 

Figure \ref{fig:eng-err} displays scalar contours on a $z-$plane that roughly bisects the advancing flame front. The first column presents the ground truth (\ghr) for $y-$velocity, temperature, and the mass fractions of key chemical species (\ce{H2}, \ce{H2O}, together with the hydroxyl (\ce{OH}) and the hydroperoxy (\ce{HO2}) radicals.) The second and third columns depict the absolute errors between the \gint~ and \gsr~ predictions, respectively, and the ground truth. Notably, \gint, derived from low-resolution data, is already well resolved given the large number of elements for the domain size, with most of its errors concentrated along the flame front. In this context, super-resolution acts as a targeted refinement, thinning these erroneous regions for the three scalars compared with \gint. This is particularly significant because reduced-order models in turbulent combustion are inherently nonlinear, rendering even minor enhancements essential for maintaining physical accuracy. 

{For completeness, we also examine the \(y\)-velocity component, which is oriented perpendicular to the piston motion. In this configuration, this component reflects both the large-scale tumble induced by the engine geometry and the flow induced by flame-front dilatation. Unlike the thermochemical scalars, for which the dominant discrepancies are localized to the reaction zone, the \(u_2\) field exhibits errors distributed throughout the region of interest, with two discernible patterns. One aligns with the flame-front topology, while the other appears as spatially scattered patches. Overall, the absolute errors remain modest, with the majority below \(10\%\). Consistent with the thermochemical fields, super-resolution reduces the error concentrated along the flame front. However, the more broadly distributed discrepancies persist. Addressing this residual component, which is not uniquely tied to the flame-front, is identified as a direction for future work.}

\begin{figure}[h!]
\setlength{\unitlength}{1cm}
\hspace{-0.5cm}
\includegraphics[width=7.5cm]{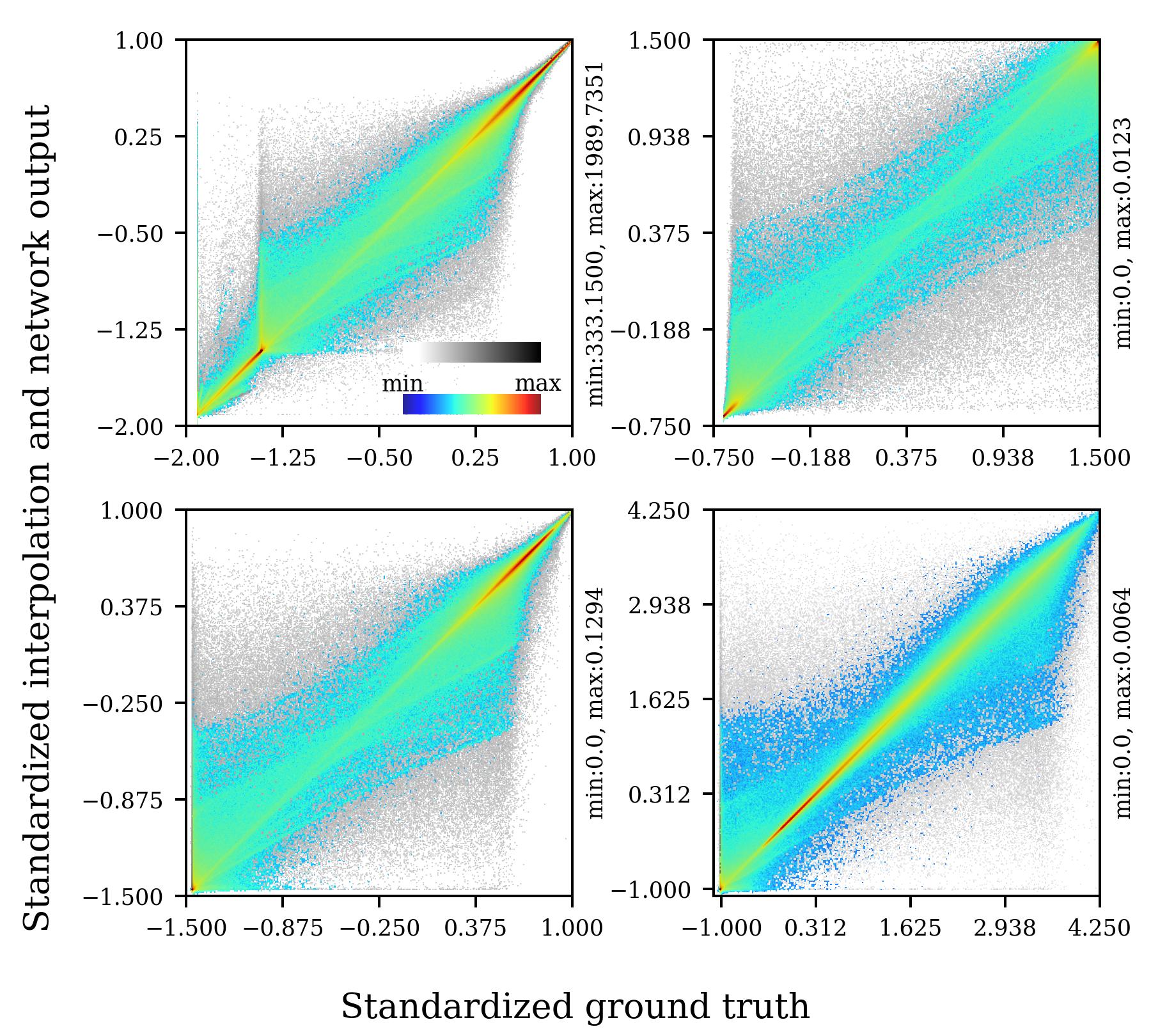}
\put(-6.0,6.1){{$T$ (K)}}
\put(-2.5,6.1){{$Y_{\ce{H2}}$}}
\put(-6.0,3){{$Y_{\ce{H2O}}$}}
\put(-2.5,3){{$Y_{\ce{OH}}$}}
\caption{\footnotesize Joint PDFs computed between interpolation/network output and ground truth for temperature and mass fractions of three species for Case 2. Grayscale color scheme is used for interpolation and the jet colormap is used for GNN output.}
\label{fig:eng-jpdfs}
\end{figure}
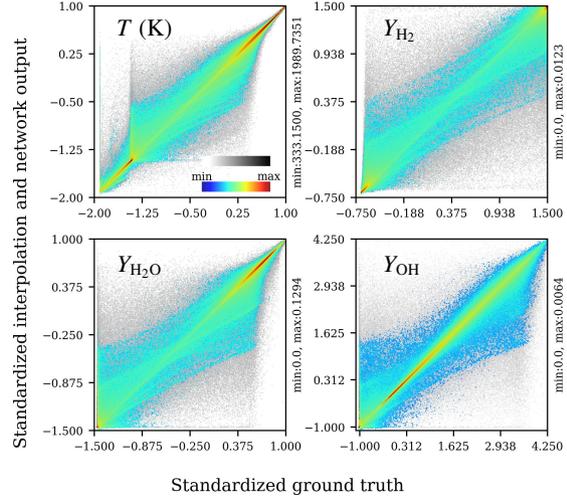

Due to memory constraints, a subdomain encompassing the flame front and the burnt region is extracted to ensure adequate sampling of nodes across all variables. Joint probability density functions (PDFs) presented in Fig. \ref{fig:eng-jpdfs} further quantify the observed discrepancies and improvements. Consistent with the trends in Fig. \ref{fig:jpdfs}, the \gint~ and \ghr~ distributions exhibit pronounced misalignment, reflected in the noisy joint distributions. 
Globally, super-resolution results in improved alignment of all scalars with the ground truth. A more detailed, scalar-specific examination of \gsr~ reveals three distinct behaviors. For temperature, both \gint~ and \ghr~ demonstrate enhanced accuracy at the extremities (corresponding to the reactant and product zones, respectively), while the intermediate region near the flame front remains challenging to reconstruct. In contrast, the mass fraction of \ce{H2} exhibits localized errors at the extremities but improved accuracy near the flame front. For the mass fractions of \ce{H2O} and \ce{OH}, errors at the peak values (product zone) are already minimal for \gint~and are well preserved by \gsr. Near these peaks, \gsr~ yields additional error reduction, whereas at low values (reactant zone), \gsr~ still displays minor residual noise despite noticeable improvement. These observations indicate that the interpolation technique exerts a measurable influence on the super-resolution performance, underscoring the necessity for interpolation-agnostic super-resolution methodologies in future investigations.

\begin{table*}[h!]
\centering
\caption{Comparison of percentage errors (\%) for IDW interpolation and super-resolution across all scalars.}
\label{tab:error_comparison_engine}

\resizebox{\textwidth}{!}{%
\begin{tabular}{@{}|l|c|c|c|c|c|c|c|c|c|c|c|c|c|@{}}
\hline
 & $u_1$ & $u_2$ & $u_3$ & $p$ & $T$ & $Y_{\ce{H2}}$ & $Y_{\ce{O2}}$ & $Y_{\ce{H2O}}$ & $Y_{\ce{H}}$ & $Y_{\ce{O}}$ & $Y_{\ce{OH}}$ & $Y_{\ce{HO2}}$ & $Y_{\ce{H2O2}}$ \\
\hline
\textbf{Interpolation} & 5.23 & 5.30 & 6.63 & 0.75 & 0.67 & 0.23 & 0.13 & 1.22 & 31.46 & 16.38 & 4.66 & 42.27 & 27.65 \\
\hline
\textbf{Super-resolution} & 4.38 & 4.46 & 5.35 & 0.62 & 0.54 & 0.19 & 0.10 & 0.96 & 25.36 & 12.88 & 3.64 & 34.27 & 22.44 \\
\hline
\textbf{Reduction (\%)} & 16.25 & 15.85 & 19.31 & 17.33 & 19.40 & 17.39 & 23.08 & 21.31 & 19.39 & 21.37 & 21.89 & 18.93 & 18.84 \\
\hline
\end{tabular}%
}
\end{table*}

Lastly, the cumulative error is computed following Equation (\ref{eqn:cumulative-err}), {and presented in Table \ref{tab:error_comparison_engine}}. It is worth reiterating that, owing to the large number of elements, the low-resolution fields and consequently the interpolated counterparts are already well resolved, although not at the DNS level. Among the hydrodynamic scalars, except for pressure, the errors are minimal. Temperature, as well as the reactant and product species, exhibit even lower errors, whereas intermediate species display noticeably higher discrepancies. For each thermochemical scalar, regions away from the flame front correspond to low variability. The GNN-based super-resolution model, consisting of collections of multilayer perceptrons, can introduce localized noise due to its inherent nonlinearity. Nevertheless, the overall error across all thermochemical scalars is consistently reduced. On average, the error incurred by \gsr~ is approximately 20\% lower than that of \gint.

\section{Conclusions\label{sec:conclusions}}

We present a novel machine learning methodology based on graph neural networks (GNNs) for super-resolving turbulent reacting flows on complex meshes. The interpolation-free approach in conjunction with edge-node convolutions allows for the accurate and systematic reconstruction of important flow features such as gradients. The GNN-based approach operates directly on the native non-uniform or unstructured mesh, eliminating the need for intermediate interpolation. The implementation on state-of-the-art GPUs enables application on large amounts of data and practically relevant applications. 

The results based on the reacting channel and hydrogen engine datasets show significant advantages of the proposed approach compared to simpler approaches such as interpolation, especially for statistics relevant to modeling, such as joint PDFs. This enables a wide range of future application scenarios for the method presented here, from experimental investigations of laboratory flames to the development of industrially relevant setups. While these results demonstrate the promise of the approach, a more fundamental analysis aimed at understanding model behavior, such as identifying conditions under which the method may succeed or fail will be pursued in future studies involving lower dimensionality or fewer degrees of freedom.

While training and application were conducted on state-of-the-art hardware, computational performance was not the primary focus of this work. Optimization strategies are being explored to suit exascale supercomputers, in a bid to leverage their immense potential for supporting advancements in green energy technologies. In the future, a detailed investigation will be conducted to assess how multi-modality can be better integrated into the proposed approach to expand its applicability.

\section*{CrediT authorship contribution statement}
\textbf{Priyabrat Dash}: conceptualization, methodology, software, formal analysis, visualization, writing (original draft). 
\textbf{Konduri Aditya}: conceptualization, methodology, supervision, writing (review and editing)
\textbf{Christos E. Frouzakis}: conceptualization, data curation, supervision, visualization, writing (review and editing)
\textbf{Mathis Bode}: conceptualization, methodology, data curation, supervision, funding acquisition, writing (review and editing)

\section*{Declaration of competing interest} 
The authors declare that they have no known competing financial interests or personal relationships that could have appeared to influence the work reported in this paper.

\section*{Acknowledgments}
The authors gratefully acknowledge the computing time granted by the John von Neumann Institute for Computing (NIC) and provided on the supercomputer JURECA at Jülich Supercomputing Centre (JSC). Furthermore, access to JEDI via the JUPITER Research and Early Access Program (JUREAP) is acknowledged. JEDI was a smaller test system for JUPITER that has received funding from the European High Performance Computing Joint Undertaking (JU) as well as from the German Federal Ministry of Research, Technology and Space (BMFTR) and the Ministry of Culture and Science of North Rhine-Westphalia (MKW NRW). PD acknowledges the support from the Prime Minister's Research Fellowship, India and the Helmholtz Visiting Research Fellowship. KA is supported by the ANRF Core Research Grant, India.

\section*{Supplementary material}

\begin{figure}[t]
\centering
\vspace{-0.3 cm}
\includegraphics[width=7cm]{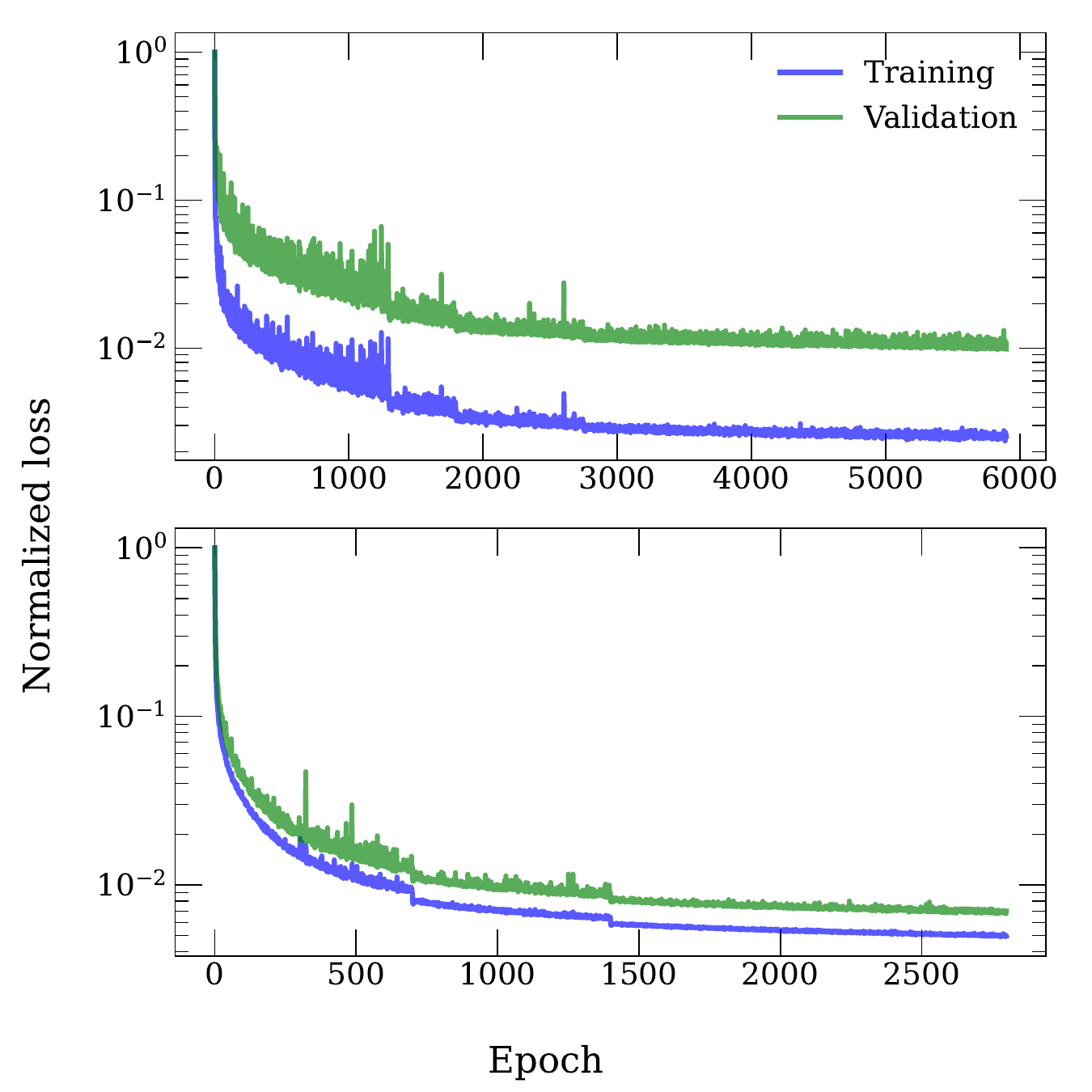}
\centering
\caption{\footnotesize Normalized loss curves for reacting channel (Case 1, top) and reacting IC engine (Case 2, bottom).}
\label{fig:losscurve}
\end{figure}

{While mean squared error (MSE) is a standard metric for assessing the convergence of ML training, it is dominated by large-scale structures and may not reliably capture the quality of small-scale reconstruction. Nonetheless, we have plotted the loss curves for both cases in Fig. \ref{fig:losscurve}.}

\FloatBarrier

\bibliographystyle{cnf-num}
\bibliography{main}

\end{document}